\documentclass[a4paper,twocolumn,11pt,accepted=2025-12-21]{quantumarticle}
\pdfoutput=1
\usepackage[utf8]{inputenc}
\usepackage[english]{babel}
\usepackage[T1]{fontenc}
\usepackage{amsmath}
\usepackage{hyperref}
\usepackage{tikz}
\usepackage{mathtools}
\usepackage{latexsym}
\usepackage{graphicx}
\usepackage{float}
\usepackage{dcolumn}
\usepackage{bm}
\usepackage{amssymb}
\usepackage{mathtools}
\usepackage[space]{grffile}
\usepackage{xcolor}
\usepackage{comment}
\usepackage{cancel}
\usepackage[export]{adjustbox}
\usepackage{soul}

\begin{document}
\title{Isolated zero mode in a quantum computer from a duality twist}
\author{Sutapa Samanta}
\affiliation{Department of Physics and Astronomy, Western Washington University, Bellingham, Washington 98225, USA}
\orcid{0000-0002-1872-1864}
\author{Derek S. Wang}
\affiliation{IBM Quantum, IBM T.J. Watson Research Center, Yorktown Heights, New York 10598, USA}
\author{Armin Rahmani}
\affiliation{Department of Physics and Astronomy, Western Washington University, Bellingham, Washington 98225, USA}
\affiliation{Advanced Materials Science and Engineering Center, Western Washington University, Bellingham, Washington 98225, USA}
\author{Aditi Mitra}
\affiliation{Center for Quantum Phenomena, Department of Physics, New York University, 726 Broadway, New York, New York 10003, USA}
\orcid{0000-0002-2169-3796}

\maketitle

\begin{abstract}
Investigating the interplay of dualities, generalized symmetries, and topological defects beyond theoretical models is an important challenge in condensed matter physics and quantum materials. A simple model exhibiting this physics is the transverse-field Ising model, which can host a topological defect that performs the Kramers-Wannier duality transformation. When acting on one point in space, this duality defect imposes the duality twisted boundary condition and binds a single zero mode. This zero mode is unusual as it lacks a localized partner in the same $\mathbb{Z}_2$ sector and has an infinite lifetime, even in finite systems. Using Floquet driving of a closed Ising chain with a duality defect, we generate this zero mode in a digital quantum computer. We detect the mode by measuring its associated persistent autocorrelation function using an efficient sampling protocol and a compound strategy for error mitigation. We also show that the zero mode resides at the domain wall between two regions related by a Kramers-Wannier duality transformation. Finally, we highlight the robustness of the isolated zero mode to integrability- and symmetry-breaking perturbations. Our findings provide a method for exploring exotic topological defects, associated with noninvertible generalized symmetries, in digitized quantum devices.
\end{abstract}

\section{Introduction}
Symmetry is a cornerstone of modern physics. Traditionally, symmetries in quantum systems are realized as unitary representations of a group of transformations that leave the Hamiltonian invariant. Dualities are transformations that relate two different models, or different phases of the same model, and are closely related to symmetry but cannot always be described within the traditional framework~\cite{Cobanera2011,SEIBERG2016395,Nahum21,Verstraete21}. 
Duality transformations can act as projectors, which do not admit a unitary representation but are examples of noninvertible symmetries~\cite{VERLINDE1988360,Bhardwaj,Fendley16,Thorngren19,Fendley20}. Both unitary (invertible) and noninvertible symmetries are described within the recently developed framework of \textit {generalized symmetries}~\cite{Wen20,McGreevy22}, and can manifest through topological defects~\cite{Gaiotto2015,Fendley16,Fendley20,Gaiotto_2021}. 

Exploring topological defects, encompassing both invertible and noninvertible types, stands at the forefront of research in condensed matter and high-energy physics \cite{cheng2023liebschultzmattis,eck2023xxz}. Despite remarkable theoretical progress, detecting exotic topological defects associated with generalized symmetries is challenging in traditional condensed-matter systems, rendering their experimental realization an open question. 
Here, 
we realize such a system on a synthetic platform, a small, noisy quantum computer~\cite{Preskill2018}. 

The Kramers-Wannier (KW) duality transformation of the transverse-field Ising model (TFIM) is one of the most well-known dualities \cite{Kramers1941}. The TFIM can support different topological defects, including the familiar unitary spin-flip defect, which captures the $\mathbb{Z}_2$ symmetry of the model, and the duality defect \cite{Fendley16,Fendley20,Tan22} studied here.The duality defect is one of the simplest examples of a noninvertible transformation as it projects out states of a given $\mathbb{Z}_2$ symmetry \cite{Fendley16}. Recently, the integrability of the critical TFIM was studied in Ref. \cite{sinha2025} (see also Ref.~\cite{Zhang2025} for a wave-function perspective) and many conserved charges were found, including charges associated with the KW duality and other noninvertible symmetries. A duality defect acting at a particular point in space imposes a special ``twisted'' boundary condition \cite{CARDY1989581,VERLINDE1988360,Schutz_1993,Oshikawa97,Petkova00,Grimm:2001dr,FrohlichFucksRunkelSchweigert,Chang:2018iay}. One of the consequences of the duality defect is the emergence of a zero mode, bound to the defect, with distinct characteristics from more well-known edge zero modes occurring in the TFIM with open boundary conditions~\cite{Fendley16,Tan22}.

It is well-established that the TFIM with open boundary conditions can host Majorana zero modes \cite{Kitaev01, FendleyXYZ} in the topological phase, both in equilibrium and in its Floquet versions \cite{Sen13,Yates19}. These modes appear in pairs localized at the two edges of the system, and have a lifetime which grows exponentially with the system size because
they hybridize leading to an energy splitting that is exponentially small in the system size. On the other hand, the zero mode appearing due to a duality twist defect is isolated, lacking a localized partner in the same $\mathbb{Z}_2$ sector. It thus has an infinite lifetime, even in finite systems.  In addition, this isolated mode exists for all couplings of the static and Floquet versions of the TFIM, and separates the chain into two parts whose couplings are related by the Kramers-Wannier duality transformation \cite{Fendley16,mitra2023nonintegrable}.

Digitized quantum devices have provided a new playground for realizing topological physics~\cite{Rahmani2020,Satzinger2021,Stenger2021,Liu2022,Kirmani2022,Andersen2023,Shtanko2023,Kirmani2023}. In particular, Floquet versions of the TFIM with open boundary conditions have been studied on current noisy quantum devices \cite{Abanin-Aleiner22,Shtanko2023}. In this paper, we probe the Floquet version of the TFIM with a duality twist defect. We implement the duality twist defect in a closed chain of 20 qubits on the \texttt{ibmq\_kolkata} device and observe signatures of an isolated zero mode.
By obtaining the autocorrelation functions of several operators in the quantum device, we demonstrate that the zero mode, with a theoretically infinite lifetime \cite{Tan22,mitra2023nonintegrable}, is persistent. We also confirm that weak perturbations, which break both the integrability and the $\mathbb{Z}_2$ symmetry of the system, do not destabilize the zero mode in finite-length chains.

We further investigate the properties of this mode by applying a local unitary circuit that moves the twist defect. We observe that the zero mode does not follow the defect but remains in its original position. This contrasts sharply with a potential well that binds a localized mode, where moving the potential well would also translate the localized mode. This behavior is a consequence of the fact that the twist defect is topological, meaning that a local unitary transformation can move the defect, while leaving the Hamiltonian invariant when all the couplings are set equal to each other. For unequal couplings, there are still ramifications of the topological properties of the defect, which will be discussed later.  Recall that translations do not fall under the category of local unitary transformations, thus this property is not shared by a chain with a local impurity potential.

The paper is organized as follows. In Section \ref{Sec2}, we introduce the model and define the autocorrelation function utilized for detecting the zero mode. Sections \ref{Sec3}, \ref{Sec4}, and \ref{Sec5} are focused on results obtained on the IBM quantum computer. In particular, Sec. \ref{Sec3} discusses the localized zero mode for the integrable model, 
Sec.~\ref{Sec4} demonstrates the topological nature of the duality twist and its relation to the Kramers-Wannier duality transformation, Sec.~\ref{Sec5}
explores the effects of integrability- and symmetry-breaking perturbations, and Sec.~\ref{Sec6} summarizes our findings. 
Additional details are provided in three appendices.

\section{Unpaired zero mode in the Floquet-TFIM with a duality twist} \label{Sec2}
We consider the one dimensional TFIM with $L$ sites, whose Hamiltonian is given by $H =J \sum_{j=0}^{L-1} \sigma_j^z \sigma_{j+1}^z + g\sum_{j=0}^{L-1} \sigma_j^x$. The system has no open ends, and periodic boundary conditions are imposed where the operator at site $L+j$ is identified with the operator at site $j$. Under these periodic boundary conditions and spatially homogeneous couplings, there are no edge modes. One can however make the couplings spatially inhomogeneous, and in particular insert a domain wall that separates a region of $g>J$ from a region of $g<J$. Periodic boundary conditions imply that such domain walls have to appear in pairs. A zero mode resides at each domain wall, leading to a pair of zero modes for periodic boundary conditions. Open boundary conditions also lead to zero modes that appear in pairs at the two edges of the system. The modes can be viewed as Majorana operators, in a fermionic model obtained from a Jordan-Wigner transformation on the spin chain. While local in the fermionic model, these modes are nonlocal in terms of the original spin degrees of freedom. In contrast, it has been shown that duality twisted boundary conditions allow one to introduce an odd number of domain walls in the system \cite{Fendley16}, and consequently an odd number of zero modes. We discuss this scenario in more detail below.

Duality-twisted boundary conditions, e.g., on the link between sites $r,r-1$, are implemented by changing $\sigma_{r-1}^z \sigma_r^z$ to $\sigma_{r-1}^z \sigma_r^x$, and removing the transverse field $g\sigma_r^x$ \cite{Fendley16}, while still preserving periodic boundary conditions with operator $L$ identified with operator $L+j$. 
The duality twisted Hamiltonian then becomes 
\begin{align}\label{twisthamiltonian}
    H_t =H_{zz} + H_{zx} +H_x,
\end{align}
where
\begin{align*}
    &H_{zz}= J\sum_{j\ne r-1} \sigma_j^z \sigma_{j+1}^z, \,H_{zx}=J\sigma_{r-1}^z \sigma_{r}^x, \\ 
    &H_x=g\sum_{j\ne r} \sigma_j^x.
\end{align*}
 
Note that with the periodic boundary conditions, in the absence of a twist, the $\mathbb{Z}_2$ symmetry is  ${\mathcal D}_{\psi}=\prod_{j=0}^{L-1} \sigma^x_j
$. With the twist, the $\mathbb{Z}_2$ symmetry changes to $\Omega_r = i \sigma^z_r \mathcal{D}_{\psi}$.
A duality twist can also be imposed by considering an infinite chain, and gauging the $\mathbb{Z}_2$ symmetry for one half of the chain \cite{Sinha:2023hum,Seiberg:2024gek,ParayilMana:2024txy,Chatterjee:2024ych,Lu:2024ytl,Cao:2024qjj} or by performing a Kramers-Wannier transformation on one half of the chain \cite{Fei24}. One can also derive the above duality twisted Hamiltonian by using techniques from the Ising fusion category \cite{Fei24}. 

We now perform a Jordan-Wigner (JW) transformation where the JW string starts from the twist, and in particular from the site that has no transverse field. The $2L$ Majorana fermions are related to the Pauli spins as follows,
\begin{subequations}
\begin{align}
  &\gamma_{2j} =\sigma^z_{j}\prod_{k=r}^{j-1}\sigma^x_k,\,\,\\ 
&\gamma_{2j+1}=-\sigma^y_{j}\prod_{k=r}^{j-1}\sigma^x_k, \,\,\\ 
&\gamma_{2r}=\sigma^z_r,\, \, \gamma_{2r+1}=-\sigma^y_r.
\end{align}
\end{subequations}
In the above equations $j \neq r, j=0\ldots r-1,r+1,\ldots L-1 $.
It is understood that as the Majorana index $j$ increases, the JW string starts from site $r$, reaches $L-1$ and then continues to site $0,1 \ldots$. In particular $\gamma_{2r-1}= -\sigma^y_{r-1}{\mathcal D}_\psi \sigma^x_{r-1}$. We will use this below when writing the twist term $H_{zx}$ in terms of Majorana fermions.

With the twist, the $\mathbb{Z}_2$ symmetry in terms of the Majorana fermions is
\begin{align}
&\Omega_r = i \sigma^z_r \mathcal{D}_{\psi}=(-i)^{L-1}\biggl[\gamma_{0}\gamma_{1}
\gamma_{2}\gamma_{3}\ldots\nonumber\\
&\gamma_{2r-2}\gamma_{2r-1}\gamma_{2r+1}\gamma_{2r+2}\gamma_{2r+3}\ldots \gamma_{2L-2}\gamma_{2L-1}\biggr].
\end{align}
Note that $\Omega_r$ is a product of $2L-1$ Majorana fermions as it
does not contain the $\gamma_{2r}$ Majorana fermion. Since it is a
string containing odd number of Majorana fermions, we have the following relations
\begin{align}
    &\Omega_r \gamma_{2r} = -\gamma_{2r}\Omega_r,\\
    &\Omega_r \gamma_{j} = \gamma_{j}\Omega_r, \hspace{1.65cm} j\neq 2r
\end{align}
The Hamiltonian $H_t$ in terms of the Majorana fermions becomes
\begin{align}
&H_{zz}= -i J\sum_{j=0\ldots L-1,j\neq r}\gamma_{2j-1}\gamma_{2j},\\ 
&H_{x}= -i g\sum_{j=0\ldots L-1,j\neq r}\gamma_{2j}\gamma_{2j+1},\\
&H_{zx}= -i J \Omega_r\gamma_{2r-1}\gamma_{2r+1}.
\end{align}

 Even though there are $2L$ Majorana fermions in the Hilbert space, $H_t$ explicitly contains $2L-1$ Majorana fermions as $\gamma_{2r}$ does not enter in $H_t$. Importantly, although $\gamma_{2r}$ does not appear in the Hamiltonian, it is not a zero mode as it does not commute with $\Omega_r$ and consequently the Hamiltonian. The string operator appears to make the system interacting, but the problem is still solvable because the $2L-1$ Majorana fermions entering in $H_t$ commute with $\Omega_r$. Thus one may set $\Omega_r=\pm 1$, after which the Hamiltonian is a bilinear in the Majorana fermions. Thus $H_t$, has eigenmodes obtained by diagonalizing a $(2L-1)\times (2L-1)$ matrix for each $\mathbb{Z}_2$ sector. Since the matrix dimension is odd in each $\mathbb{Z}_2$ sector, this has to yield an odd number of zero modes in each $\mathbb{Z}_2$ sector. 

For the parameters above, a given $\mathbb{Z}_2$ sector hosts a  zero mode that is localized at $r$, and has a localization length $\xi$, where $\exp[1/\xi]\approx \text{max}(g,J)/\text{min}(g,J)$ \cite{Tan22}.
There is nothing special about the sites $r,r-1$, and a local unitary transformation can move the twist defect from one pair of sites to the 
neighboring pair of sites \cite{Vidal16,Fendley16,mitra2023nonintegrable} without moving the zero mode \cite{mitra2023nonintegrable}. In particular, after a JW transformation this zero mode is a Majorana mode, with the branch-cut or string associated with the zero mode terminating at the duality defect \cite{Fendley16}. The local unitary transformation just modifies the length of this branch cut as it does not move the zero mode while moving the site of the duality defect.

Some approaches in the literature (see \cite{Rogerson_2022, Saleur22} and appendix of \cite{Fendley16}) have also considered a related model that is more physical for fermions, where the string operator is simply dropped to obtain the new model. In this case, there are indeed two zero modes 
with the second zero mode being 
$\gamma_{2r}$ which now commutes with the Hamiltonian (as the string operator does not explicitly enter in the Hamiltonian). However, this is a sleight of hand, which is strictly not valid if one is discussing the physics of the original spin chain. In other words, the model with two zero modes where one of them is $\gamma_{2r}$, is a different model. For the spin chain, a proper treatment always leads to the string operator $\Omega_r$, and there is no good reason to drop it. In this case $\gamma_{2r}$ is not a zero mode as it does not commute with $H_t$ due to the string $\Omega_r$.

In summary, the spin chain, has one zero mode in each $\mathbb{Z}_2$ sector, thus when looking at both even and odd $\mathbb{Z}_2$ sectors, $\Omega_r=\pm1$, there are two zero modes. In what follows, we will perform simulations that do not mix $\mathbb{Z}_2$ sectors. In this case, only a single zero mode will manifest. 
The above method of solving the problem in each $\mathbb{Z}_2 $ sector separately appears in other contexts
such as the periodic chain. For this case, the JW transformation to fermions gives a nonlocal term on one bond where the nonlocal term corresponds to fermion parity that serves as the $\mathbb{Z}_2$ symmetry of the model. Thus, in each symmetry sector, the nonlocal term is replaced by its $\pm 1$ eigenvalue, making the model local.

We now turn to the Floquet version of the above model, which is also the one that will be simulated on the IBM devices. The Floquet unitary operator for the twisted model can be written as \cite{Tan22}
\begin{equation}\label{floquetunitary}
    U =e^{-i  H_x/2} e^{-i  H_{zx}/2 } e^{ -i  H_{zz}/2}.
\end{equation}
As far as symmetries go, much of what was discussed regarding $H_t$ is also valid for $U$. In particular, in a given $\mathbb{Z}_2$ sector $\Omega_r=\pm 1$, $U$ is exactly solvable as it involves exponentials of fermion bilinears after a JW transformation. In addition, only
$2L-1$ Majorana fermions appear in $U$, so that we need to diagonalize a $(2L-1) \times (2L-1)$ dimensional matrix. Since the orthogonal matrix that evolves the $2L-1$ Majoranas only performs proper rotations, its determinant  has to be $1$. This forces an odd number of eigenvalues to be equal to $1$, 
where the corresponding eigenvectors are the zero modes.  

The analytic expression for the zero mode for the above Floquet set-up, i.e., with a duality-twist, was derived in \cite{Tan22,mitra2023nonintegrable}.  Similar to edge modes with open boundary conditions, the  zero mode with a duality twist is a linear superposition of Majoranas all along the chain. For the zero mode with the duality twist, the largest weight in the superposition is with the Majorana in the vicinity of twist. In contrast, the zero mode pairs with open boundary conditions have their largest weight at the boundaries. In addition, one may introduce a single domain wall in the duality-twisted Floquet set-up by making the couplings $g,J$
inhomogeneous \cite{Tan22}. In this case, the zero mode resides at the domain wall. As
we shall discuss further, a series of local unitary transformations can relate the homogeneous Floquet unitary of \eqref{floquetunitary} to one where a single domain wall is present \cite{mitra2023nonintegrable}. 

The Majorana zero mode with a duality twist has important differences with the Majorana pairs that appear for open
boundary conditions. While the latter occurs only in the topological phase $J>g$, the Majorana zero mode with the duality twist occurs for all values of $g,J$, including $g>J,g<J$ and even disordered ones. Only its localization length is determined by the strength of the disorder and average values of $g$ and $J$. Since the zero mode with the duality twist lacks a partner, it never decays in time, i.e., it has an 
infinite lifetime rather than a lifetime that is exponentially large in the system size. The mathematical reason for this is that the zero mode with the duality twist is a symmetry of the system as it commutes both with the generator of time evolution $U$, as well as the $\mathbb{Z}_2$ symmetry $\Omega_r$. In contrast, the Majorana pairs with open boundary conditions anticommute with the associated $\mathbb{Z}_2$ symmetry. We emphasize that a single zero mode in a given $\mathbb{Z}_2$ sector for the duality twisted model follows from a simple counting argument as summarized above.

In the spin representation, the zero mode at site $r$ has an overlap with the Pauli operator $\sigma^y_r$ \cite{Tan22,mitra2023nonintegrable}.
(In the figures, we denote $\sigma^{x,y,z}_i$ with $X_i,Y_i,Z_i$ for notational brevity.) In addition, the farther away a Majorana is from the duality twist at $r$,  the longer is the corresponding Pauli string (this is because, in order to make the model have only local terms, the JW transformation has to start from the twist).
To detect the Majorana zero mode, we employ the infinite temperature autocorrelation function, defined as
\begin{equation}\label{autocorrelation}
    A_\mathcal{O}(n) = \frac1{2^L} \text{Tr}\biggl[\mathcal{O}(n) \mathcal{O}\biggr]\ ,
\end{equation}
of an operator $\mathcal{O}$, where the Heisenberg operator after $n$ Floquet steps is given by $\mathcal{O}(n)=\left(U^\dagger\right)^n\mathcal{O}U^n$. 
The signature of the zero mode is the temporal persistence of a nonzero autocorrelation function for $\sigma^y_r$. In contrast, the autocorrelation function of other operators, with no overlap with a conserved quantity, are expected to decay to zero due to dephasing. 

To measure the autocorrelation function, we represent the trace as a sum of expectation values $\langle k|\mathcal{O}(n) \mathcal{O}|k\rangle$ in a basis chosen to be the eigenbasis of $\mathcal{O}$, $O|k\rangle=\lambda_k|k\rangle$. Each expectation value in the trace can then be measured by measuring $\langle \mathcal{O}(n)\rangle $ at time $n$ after applying the Floquet circuit to the initial state $|k\rangle$. For operators $\mathcal{O}$ corresponding to Pauli operators (or Pauli strings), the states $|k\rangle$ can be chosen to be easy-to-prepare direct products of single-qubit states. The detection of the trace can nevertheless seem challenging because we have an exponentially large number of basis states $|k\rangle$. However, a random sampling of a small number of basis states and performing a partial sum yield accurate approximations to the trace \eqref{autocorrelation}, as discussed in Appendix \ref{AppA}. 

\begin{figure*}[ht]
    \centering
    \includegraphics[width=0.8\textwidth]{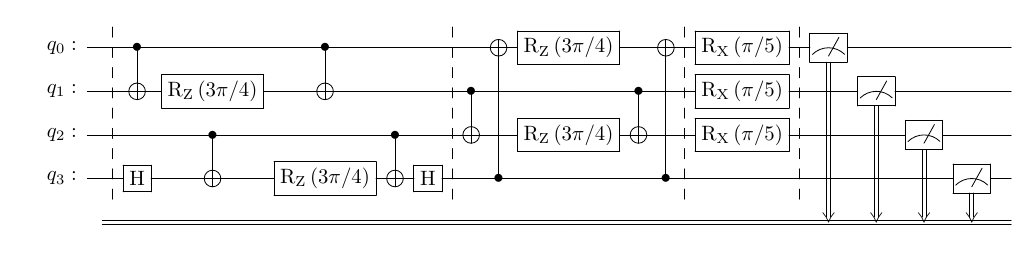}
    \caption{Implementation of the unitary \eqref{floquetunitary} for a 4-qubit system in a quantum computer. The duality twist corresponds to modifying the Floquet TFIM by removing the transverse field on qubit $3$ and applying a twist exchange interaction $\sigma^z_2\sigma^x_3$ between qubits $2$ and $3$. Periodic boundary conditions have been imposed with a $\sigma^z_3\sigma^z_0$ exchange interaction.}
    \label{fig:circuitdrawing}
\end{figure*}
\begin{figure}[ht]
    \centering
    \includegraphics[width=0.48\textwidth]{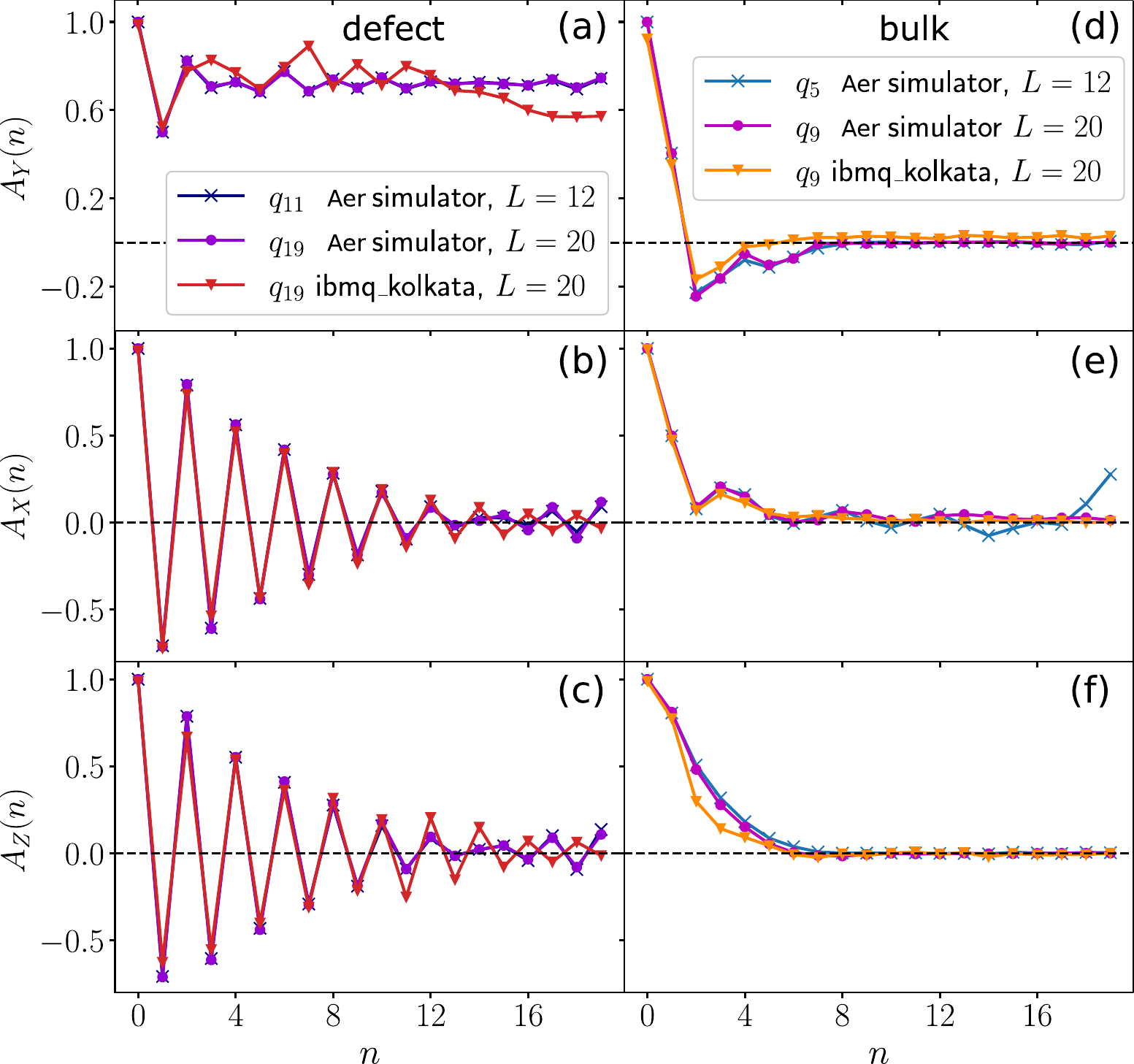}
    \caption{
    Signature of the zero mode at the defect site for the duality-twisted Floquet-TFIM. The label $q_i$ denotes the measurement at the $i$th qubit. Panel (a) shows the nondecaying autocorrelation for $\sigma^y_r$ at the defect site $r=L-1$, which overlaps with the zero mode. All other autocorrelations both at the twist defect $r=L-1$ [panel (b), (c)] and at a bulk site $j=-1+ L/2$ [panel (d), (e), (f)], far away from the twist defect, decay with stroboscopic time $n$. The chosen bulk qubit behaves qualitatively similarly to other  bulk qubits. We present data from  \texttt{ibmq\_kolkata} device for $L=20$, the longest physical loop of qubits available in the device. There is good qualitative agreement between noiseless simulation and measurements on the noisy quantum device. The noiseless simulations were also performed for a smaller system of 12 qubits. The agreement between the two system sizes $L=12,20$  indicates that we are capturing the behavior in the  thermodynamic limit. Date of demonstration at \texttt{ibmq\_kolkata} device is 18th April, 2023.
    \label{fig:XYZautocorrelation}}
\end{figure}
\section{Detection of the unpaired zero mode in the IBM quantum device}\label{Sec3}
We have chosen the values of the coupling constants to be $J=0.75\pi$ and $g=0.2\pi$ for all the computations in this paper, although similar results can be obtained for other coupling constants. To create and detect the zero mode in the quantum computer, we apply a quantum circuit that generates the Floquet dynamics by the repeated applications of a circuit block corresponding to the unitary operator in \eqref{floquetunitary}. A simple 4-qubit  version of this block, corresponding to one Trotter step, is shown in Fig.~\ref{fig:circuitdrawing}. We compute the autocorrelation~\eqref{autocorrelation} for $\mathcal{O} = \sigma^x, \sigma^y$, and $\sigma^z$ of a 20-qubit system. We perform this calculation on the 27-qubit \texttt{ibmq\_kolkata} device with a 20-qubit closed loop. Thus, the device geometry eliminates the need for using swap gates to implement periodic boundary conditions. We used a compound error mitigation method as discussed in Appendix \ref{AppC} and described in Refs.~\cite{Shtanko2023, Kirmani2023}.

As mentioned above, for a Pauli operator $\mathcal{O}$, we need to create initial states that are eigenstates of the corresponding Pauli operator.
Measuring $\mathcal{O}=\sigma^z$ is the simplest as the input and output in IBM devices are measured in the eigenbasis of $\sigma^z$ by default. However, both initialization and measurement in a different basis can be implemented through rotation operators $R_{y}(\pi/2)$ ($R_x(-\pi/2)$) that rotate from the $z$ to the $x$ ($y$) direction.

In Fig.~\ref{fig:XYZautocorrelation}, we present both noiseless simulation and noisy quantum-hardware results for the bulk and defect autocorrelation functions obtained from the partial trace with 20 random basis states.
We also show noiseless simulation results from Qiskit Aer simulators for 12-qubit and 20-qubit systems. The simulation results exhibit very small system-size dependence, indicating that the 20-qubit system accurately captures the thermodynamic limit. Furthermore, the results from the \texttt{ibmq\_kolkata} device for the 20-qubit system is in good agreement with the noiseless simulations. We observe that the autocorrelation function of $\mathcal{O}=\sigma^y_r$ at the twist defect $r=L-1$ does not decay, which overlaps with the emergent Majorana zero mode. In contrast, all other autocorrelators decay to zero in the long time limit. The persistence of the autocorrelation function of the operator corresponding to the unpaired zero mode, during the Floquet dynamics, provides evidence for the dynamical generation of this mode in a digitized quantum device.

\section{Unitary translation of twist defect}\label{Sec4}
\begin{figure}[ht]
    \centering
    \includegraphics[width=0.47\textwidth]{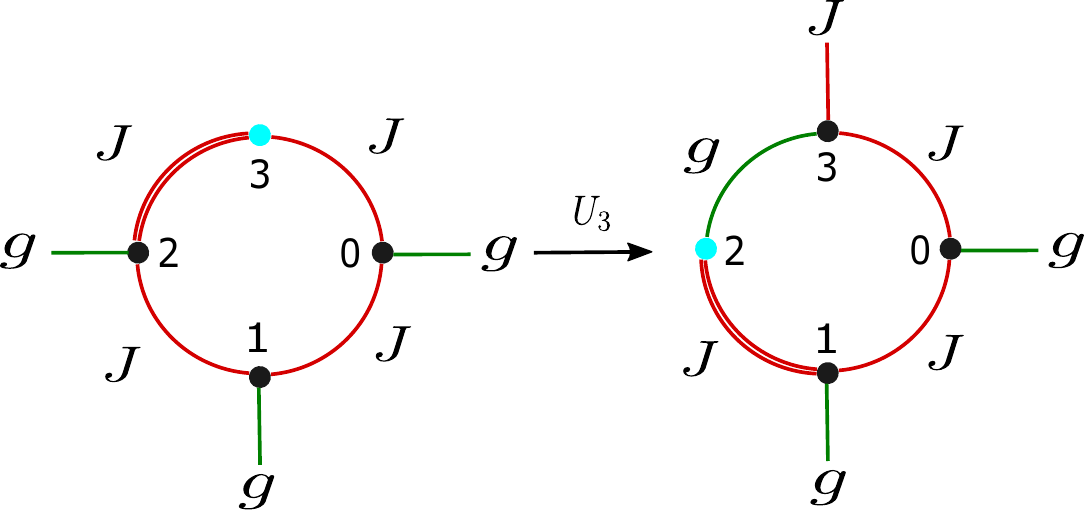}
    \caption{A twist defect that was originally on sites 2 and 3 (left) is moved to sites 1 and 2 (right) by a unitary transformation $CZ_{2,3}H_2$. 
    Cyan denotes the site $r$ at which there is no magnetic field while the double red lines denote a twist interaction $\sigma^z_{r-1}\sigma^x_r$. Vertical lines denote a transverse field while solid lines between sites $j,j+1$ denote a $\sigma^z_j\sigma^z_{j+1}$ interaction.
    After the unitary transformation (right), note the Kramers-Wannier duality between sites 2,3 and the rest of the chain where the strength of the transverse field at site 3 is $J$ while the exchange interaction between qubits 2 and 3 is $g$. 
    The unitary transformation shifts the twist but not the Majorana zero mode which stays localized at site 3. Thus the Majorana zero mode resides at the domain wall across which $g,J$ exchange roles. }
    \label{fig:TranslationDrawing}
\end{figure}
\begin{figure}[ht]
    \centering
    \includegraphics[width=0.45\textwidth]{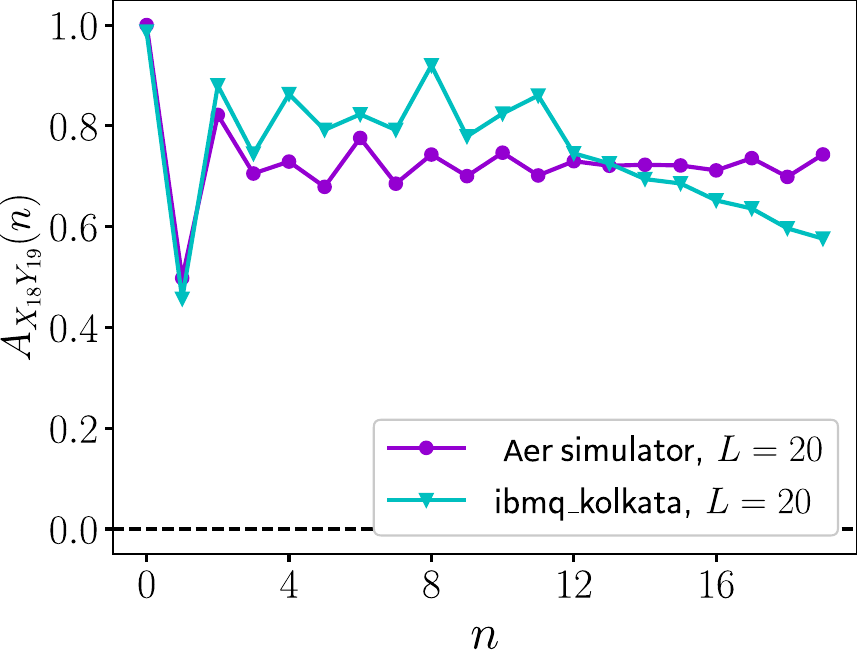}
    \caption{
    Unpaired Majorana in the \texttt{ibmq\_kolkata} device for $L=20$ after a unitary translation that has moved the twist defect from sites $18,19$ to sites $17,18$. The Majorana stays localized at site 19 but now has an overlap with the longer Pauli string $\sigma^x_{18}\sigma^y_{19}$. Date of demonstration is 5th May, 2023.}
    \label{fig:UnitaryTranslation}
\end{figure}
A hallmark of a topological defect is that it can be moved by local unitary transformations, while leaving the model invariant. Strictly speaking the model is invariant only for $g=J$, and in the high frequency limit. Nevertheless, even away from this special case, physical consequences of the topological nature of this defect manifest in the invariance of the position of the zero mode 
as the twist defect is moved.

A local unitary transformation of the form $U_{r} = CZ_{r-1, r} H_{r-1}$, where $CZ_{r-1,r}$
is a controlled-Z gate, and $H_{r-1}$ is a Hadamard gate, moves the twist-defect from the sites $(r-1,r)$ to sites $(r-2,r-1)$ \cite{Fendley16,mitra2023nonintegrable}.  Fig.~\ref{fig:TranslationDrawing} illustrates this for a 4-qubit system. The unitary $U_{r}$ introduces a transverse field at site $r$ of strength $J$ while eliminating the transverse field at site $r-1$. It also changes the terms $J\sigma_{r-2}^z \sigma_{r-1}^z$ and $J\sigma_{r-1}^z \sigma_{r}^x$ to $J\sigma_{r-2}^z \sigma_{r-1}^x$ and $g\sigma_{r-1}^z \sigma_{r}^z$. Thus a Kramers-Wannier duality transformation has been performed on the sites $r,r-1$ where the roles of $g,J$
have been interchanged. Remarkably, the Majorana zero mode stays localized at $r$, and now separates two regions of the chain that are related by the Kramers-Wannier duality transformation. Although the Majorana zero mode stays localized at site $r$, in the language of Pauli strings, it is a more nonlocal object. In particular, instead of having an overlap with $\sigma^y_r$, it now has an overlap with 
$\sigma^x_{r-1}\sigma^y_r$.

The Majorana zero mode after a translation of the twist defect from $(18,19)$ to $(17,18)$ for a 20-qubit system is shown in Fig.~\ref{fig:UnitaryTranslation}. After one translation, the Majorana is still localized at site $r=19$, but in the language of Pauli spins, this Majorana zero mode no longer has an overlap with $\sigma^y_{19}$, but rather with the longer string $\sigma^x_{18} \sigma^y_{19}$.

\section{Effects of Interactions}\label{Sec5}
\begin{figure}[ht]
    \centering
\includegraphics[width=0.48\textwidth]{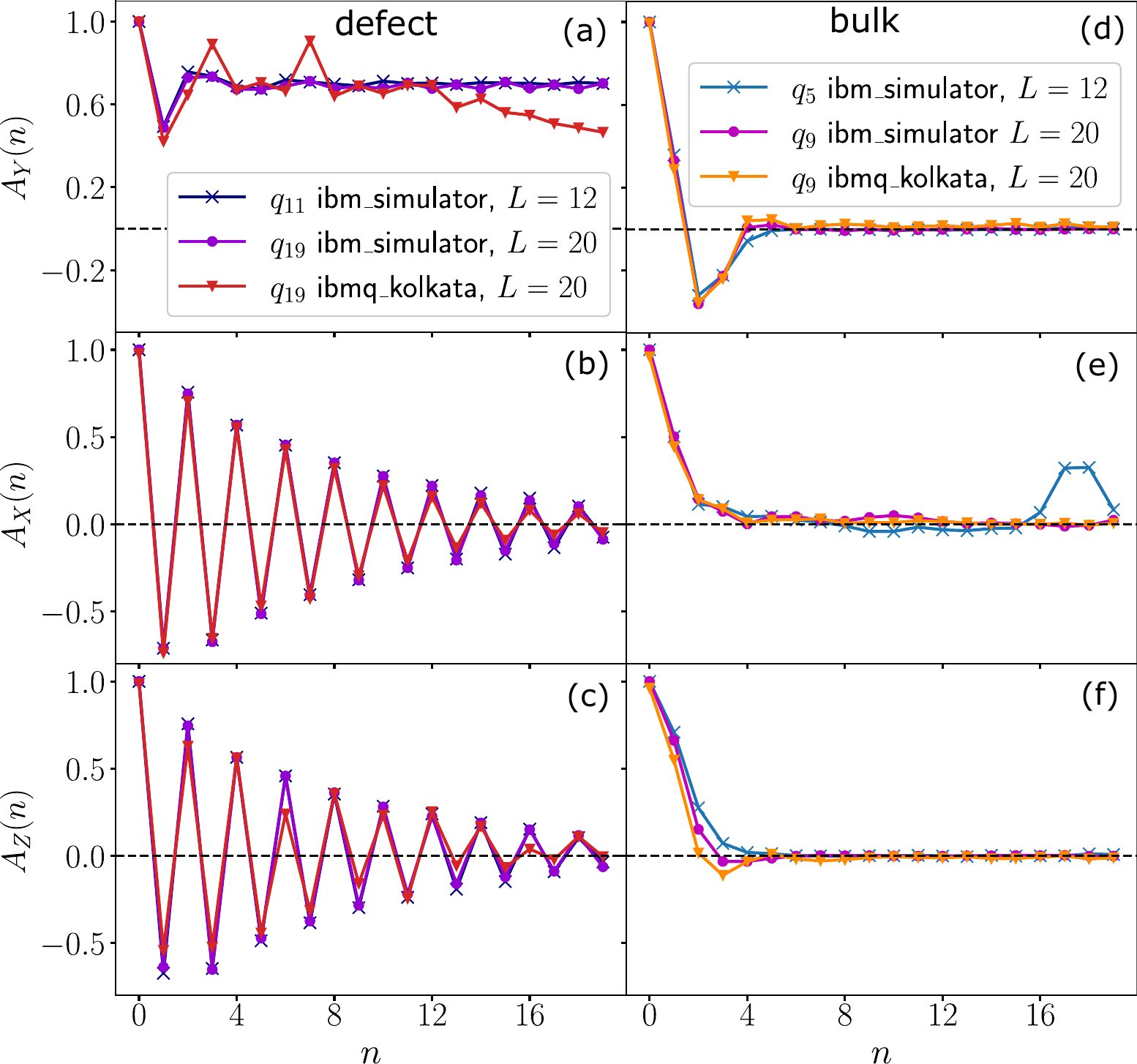}
    \caption{Signature of the Majorana zero mode at the defect site for duality-twisted interacting Floquet-TFIM. Panel (a) shows the nondecaying autocorrelation for $\sigma^y_r$ at the defect site $r=L-1$, which overlaps with the zero mode. All other autocorrelations both at the twist defect $r=L-1$ [panel (b), (c)] and at a bulk site $j=-1+ L/2$ [panel (d), (e), (f)], far away from the twist defect, decay with time. The chosen bulk qubit behaves qualitatively similarly to other bulk qubits. We present data from  \texttt{ibmq\_kolkata} device for $L=20$, the longest physical loop of qubits available in the device. There is good qualitative agreement between noiseless simulations and measurements on the noisy quantum device. The noiseless simulations were also performed for a smaller system of 12 qubits. The agreement between the two system sizes $L=12,20$  indicates that we capture the behavior in the thermodynamic limit for these times. At longer times, we expect a stronger $L$ dependence discussed in the main text. The label $q_i$ indicates measurement of the $i$th qubit. Date of demonstrations is 22nd April, 2023.}
    \label{fig:IBMKolkataIntaractions}
\end{figure}

Now we add perturbations to the model \eqref{twisthamiltonian} through the term (other examples are presented in Appendix \ref{AppB})
\begin{equation}
 H_{xx} = J_{xx} \sum_j \sigma_j^x \sigma_{j+1}^x.   
\end{equation}
In the language of Majorana fermions, the above is a four-fermion and hence interacting term. The Floquet unitary is now given by
\begin{equation}\label{interactionunitary}
 U_{\text{int}} =  e^{-i H_{xx}/2}   e^{-i H_x/2} e^{-i H_{zx}/2 } e^{ -i H_{zz}/2}.
\end{equation}
Note that $H_{xx}$ not only breaks integrability,  but it also does not commmute with the discrete symmetry $\Omega_r$ due to the
$\sigma^x_{r\pm1}\sigma^x_{r}$ terms.
We repeat the same quantum simulation for the interacting model with the defect location at site $r=L-1$. We find that the $\sigma^y_r$-correlator at the defect remains constant, while all other correlators decay with time as shown in Fig.~\ref{fig:IBMKolkataIntaractions}. We observe a similar behavior to the noninteracting case shown in Fig.~\ref{fig:XYZautocorrelation} for both bulk and defect autocorrelation functions, within the studied number of Floquet cycles. 

\begin{figure}[ht]
    \centering
    \includegraphics[width=0.48\textwidth]{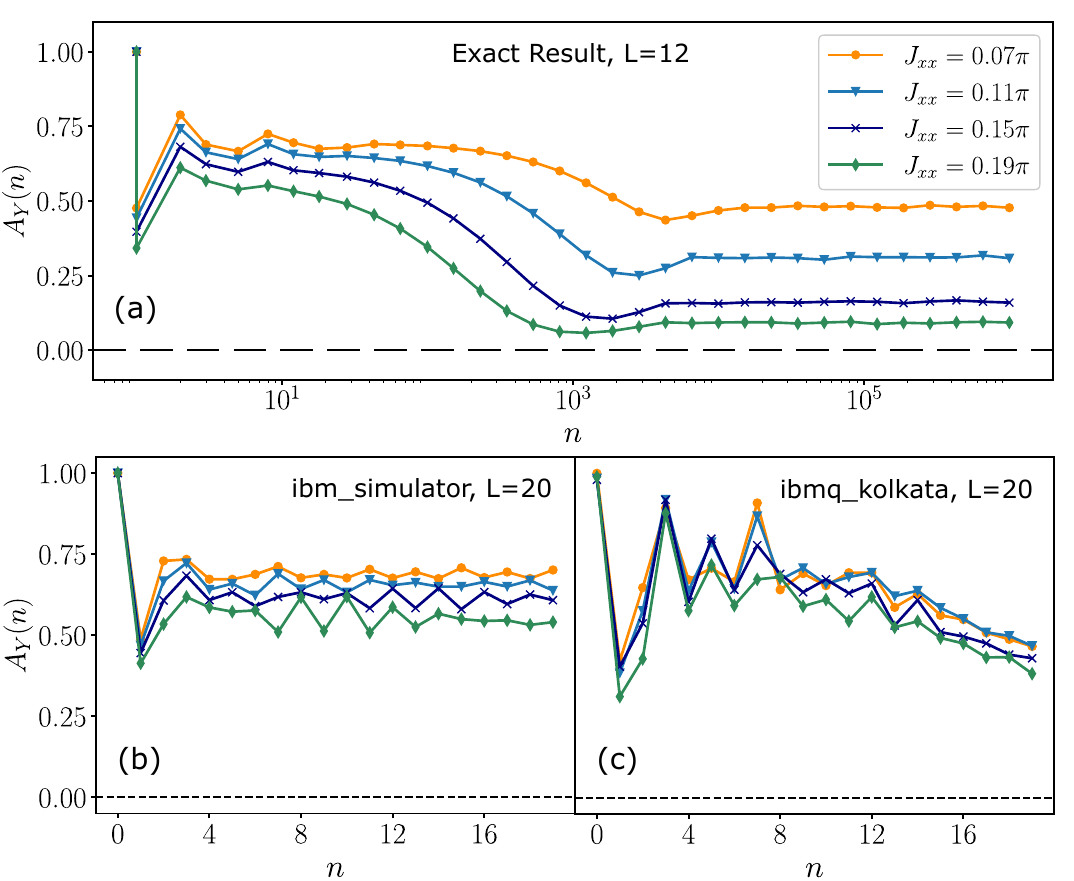}
    \caption{
    Stability of the Majorana zero mode to integrability- and symmetry-breaking perturbations in a finite system. The autocorrelation function of 
    $\sigma^y_r$ at the defect $r=L-1$, for different values of the  perturbation $J_{xx}$, are presented. Panel (a) shows the numerical results for system size $L=12$ up to large Floquet time steps. After the initial transient corresponding to the appearance of a first plateau, 
    the 
    autocorrelation stabilizes at a second plateau. This is a finite size effect where  the height of the second plateau vanishes as $L\to\infty$ \cite{mitra2023nonintegrable}. While we are not able to access this second  plateau in the long-time limit in the \texttt{ibmq\_kolkata} device due to noise, panels (b) and (c) show results from the IBM simulator and the \texttt{ibmq\_kolkata} device, respectively, for intermediate timescales.  At these timescales, the formation of the first plateau is clearly visible. Date of demonstrations is 27th April, 2023.}
    \label{fig:CompareInteractions}
\end{figure}

We note, however, that the  long time dynamics of the $\sigma^y_r$-correlator at the defect is not captured in this figure as we can run the Floquet dynamics in the device only for about 20 Floquet cycles before noise takes over. We show the theoretical behavior of the Majorana zero mode by performing exact diagonalization for long times in {Fig.~\ref{fig:CompareInteractions}(a)}. After the initial decay to the first plateau, the Majorana mode decays to a second plateau. The plateau heights, especially of the second plateau, decrease with increasing interactions $J_{xx}$. We capture only the initial portions of this behavior in the quantum device [see Fig.~\ref{fig:CompareInteractions}(b) and (c)], i.e., the decay to the first plateau. While
the first plateau height is not strongly $L$ dependent (as seen in the simulations in Fig.~\ref{fig:IBMKolkataIntaractions}),
we expect that the second  plateau height 
is $L$ dependent, where as $L\rightarrow \infty$, the plateau height decreases to zero. This is because, there is no exact zero mode when we break integrability, with the zero mode decaying via scattering with bulk excitations \cite{Yeh23}. However, a finite size system blocks these scattering processes leading to a partial decay, and hence a plateau height which is inversely related to system size \cite{mitra2023nonintegrable}. A second important observation is that while the interactions now couple the two $\mathbb{Z}_2$ sectors, yet the zero modes in the two sectors do not efficiently couple to each other as is evident by the appearance of the second plateau, which persists at long times. This is again a finite system size effect. 
Here, we introduced a specific perturbation to the model that breaks its integrability and leads to a finite decay time for the Majorana mode. In Appendix \ref{AppB}, we explore a broader class of perturbations that display similar behavior.

\section{Conclusions}\label{Sec6}
In this paper, we investigated the physics of the duality twist defect and the corresponding emergent zero mode in a periodically driven TFIM loop of 20 qubits on the IBM superconducting quantum device. The interplay of topological effects and symmetries can lead to novel phenomena. The zero mode generated by a duality twist in a TFIM chain commutes with the $\mathbb{Z}_2$ charge of the model and thus, theoretically, has an infinite lifetime in finite systems~\cite{Tan22} (in contrast to edge states with open boundary conditions that are vulnerable to exponential hybridization). Using efficient random sampling, we obtained an accurate approximation to various autocorrelation functions and provided evidence, on the quantum computer, for the emergence of the unpaired zero mode bound to the duality twist defect. Furthermore, we explored the effects of moving the duality twist defect with unitary gates, and the effect of integrability- and symmetry-breaking perturbations. This work opens a new avenue for physically creating and investigating topological defects in quantum systems, and exploring  their ramifications on nonequilibrium dynamics, a topic poorly understood theoretically. 

Future improvements in hardware quality and error mitigation may allow accessing longer evolution times with deeper circuits, providing access to the long-time plateaus of persistent autocorrelation functions in the presence of interactions.
Exciting directions of future research include implementing topological defects on the quantum computer, not only in
$Z_n$ spin chains \cite{Fendley20}, but also in two-dimensional Floquet codes \cite{Hastings2021dynamically}.

\begin{acknowledgments}
We thank Oles Shtanko for helpful comments and discussions.
We
acknowledge the use of IBM Quantum services. We also
thank the Brookhaven National Laboratory for providing access to IBM devices. This work was supported by the US National Science Foundation under Grants NSF DMR-2018358 (AM) and DMR-1945395 (AR and SS), and in part by the US National Science Foundation under Grant No. NSF PHY-1748958 (AR and AM). 
\end{acknowledgments}
\bibliographystyle{quantum}
\bibliography{biblio}

\onecolumn\newpage
\appendix

\section{Convergence of the partial trace} \label{AppA}
\begin{figure}[ht]
    \centering
    \vspace{5mm}
\includegraphics[width=0.55\textwidth]{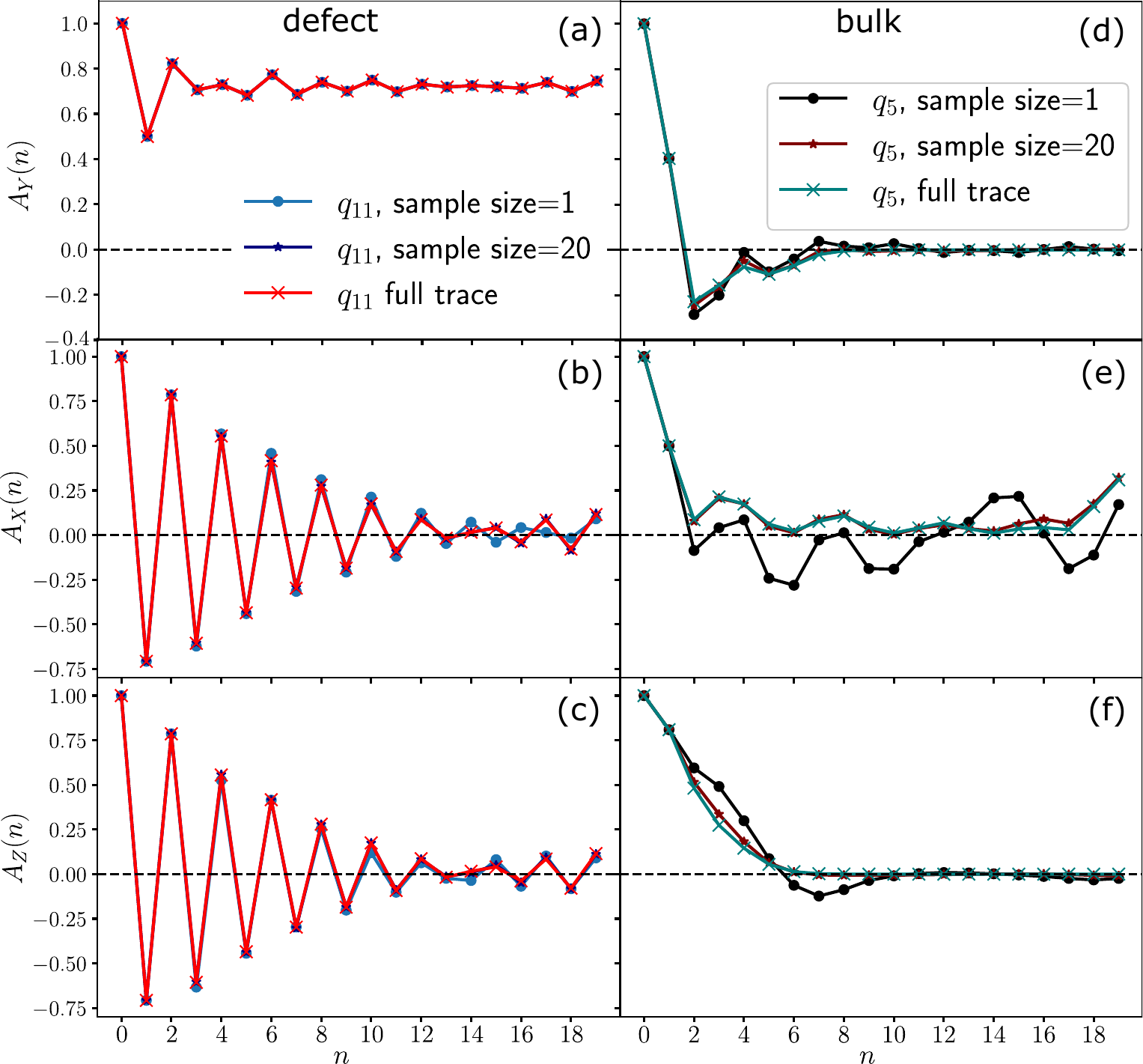}
    \caption{Autocorrelation functions from computing the partial trace in a classical computer for different sample sizes of random initial states, for system size $L=12$. The left panels [panel (a), (b), (c)] show the autocorrelation functions at the site of twist defect $r=L-1$, while the right panels [panel (d), (e), (f)] are for a site far away from it, $j=-1+L/2$.}
    \label{fig:exact}
\end{figure}

In this section, we provide evidence for the accurate approximation of the autocorrelation functions (based on the full trace) by a partial trace with a small number of random basis states.
In Fig.~\ref{fig:exact}, we show the expectation value with a single random state, and the partial trace with 20 random initial states, and we compare them with the full trace calculated with all $2^{12}$ basis states for a system of size $L=12$. We have calculated these quantities for all three Pauli operators. Even a single expectation value is relatively close to the exact (full trace) autocorrelation function, for both bulk and defect sites, due to the little variation between these expectation values. The results of the partial trace with 20 random states are in excellent agreement with the full trace, supporting the use of the partial trace to approximate the autocorrelation functions.

{All expectation values are the same, and correspond to the $Y$ autocorrelation function at the defect site. Thus, for this particular observable, the trace is independent of the number of basis states used.}


\section{Effects of other perturbations on stability}\label{AppB}
\begin{figure}[ht]
    \centering
    \includegraphics[width=0.8\linewidth]{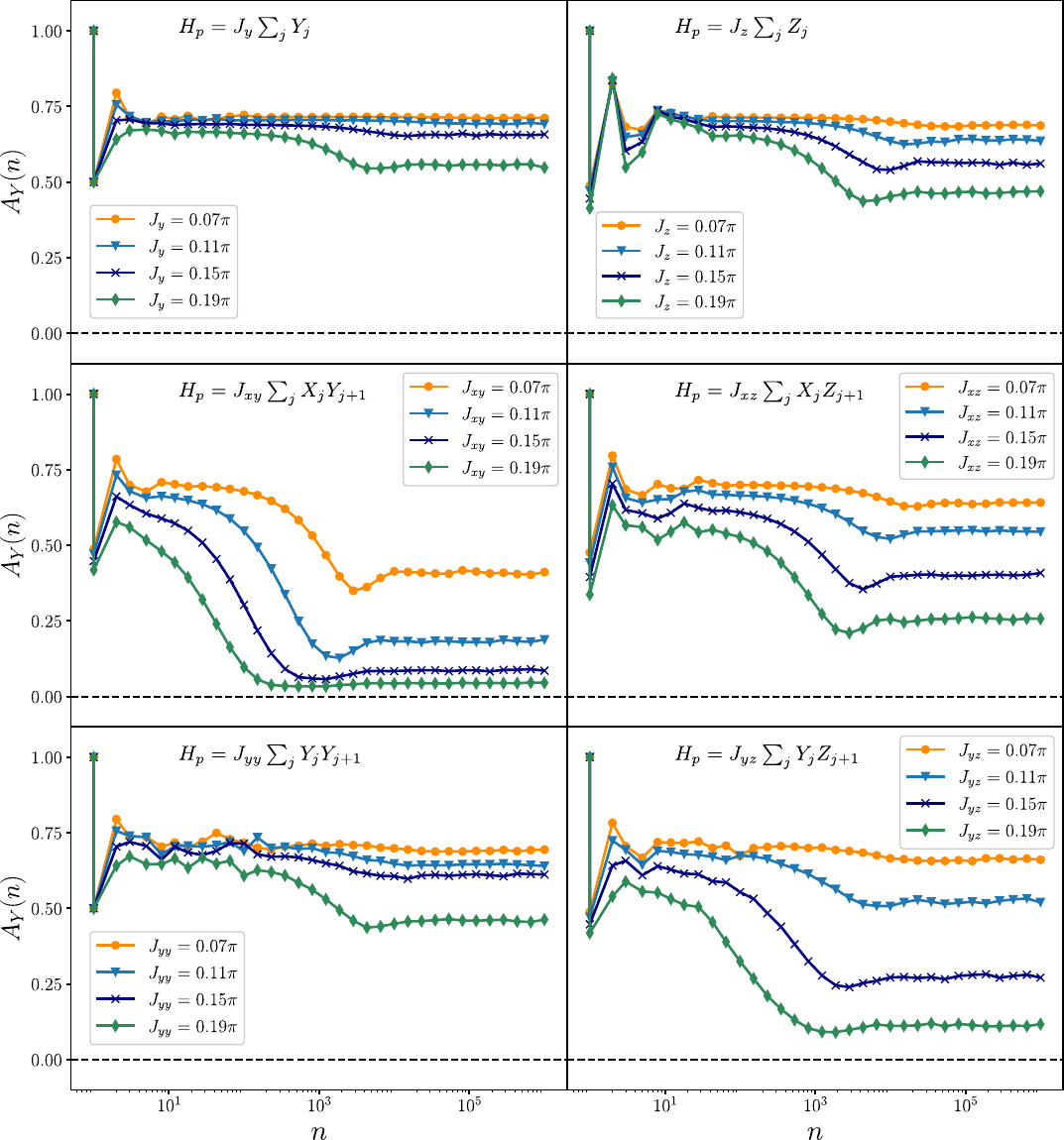}
    \caption{Stability of the Majorana zero mode to various integrability- and symmetry-breaking perturbations in a finite size system of $L=12$, obtained from exact diagonalization. 
    The autocorrelation function of 
    $\sigma^y_r$ at the defect $r=L-1$ are presented as a function of Floquet steps, for different coupling constants. The autocorrelation functions decay to a non-zero value at long-times, where the plateau heights decrease with increasing coupling constants and also with increasing system sizes (not shown). In the thermodynamic limit $L\to \infty$, all the autocorrelations are expected to decay to zero.}
    \label{fig:interact}
\end{figure}

We show the stability of zero mode in the presence of various integrability- and symmetry-breaking perturbations. We restrict ourselves to only one-qubit and two-qubit terms for simplicity. We apply perturbations of the form $H_\text{int} = \lambda\sum_j h_j$, where $h_j$ is taken to be $\sigma^y_j$, $\sigma^z_j$, $\sigma^x_j \sigma^y_{j+1}$, $\sigma^x_j\sigma^z_{j+1}$, and $\sigma^y_j\sigma^z_{j+1}$. We note that $\sigma^z$ and $\sigma^y$ are nonlocal in terms of the Majorana operators as they contain Jordan-Wigner strings.
The results for the $A_Y$ autocorrelation function are shown in Fig.~\ref {fig:interact}.

\section{Error mitigation}\label{AppC}
We employ a composite error mitigation strategy combining error suppression and mitigation techniques~\cite{Ferris:2022hvp}. Noisy bitstrings resulting from each shot are first mitigated by correcting readout errors using the method of Ref.~\cite{Nation2021M3}, which utilizes calibration circuits scaling linearly with the number of qubits. Finally, zero-noise extrapolation is applied by extrapolating measured probabilities to the zero-noise value \cite{Li2017ZNE, Temme2017PECandZNE}. This is achieved by performing measurements with 8,000 shots at noise factors of 1 (applying the original circuit) and 3 (evolving the system forward by the original circuit, backward by appending the inverse of the original circuit, and finally forward again by reapplying the original circuit \cite{Rivero2022ZNE}). A line through 1 and 3 on the horizontal axis yield the zero-noise extrapolated value at the intercept with the vertical axis.

\section{Layout of the computing device and its characteristics}
\begin{figure}[ht]
\centering
    \includegraphics[width=0.47\textwidth]{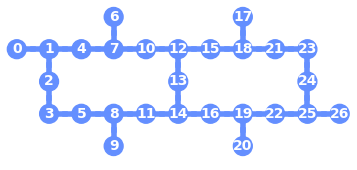}
    \caption{Qubit layout of \texttt{ibmq\_kolkata} device }
    \label{fig:kolkata_layout}
\end{figure}

\begin{figure}[]
    \centering
    \includegraphics[width=0.48\textwidth]{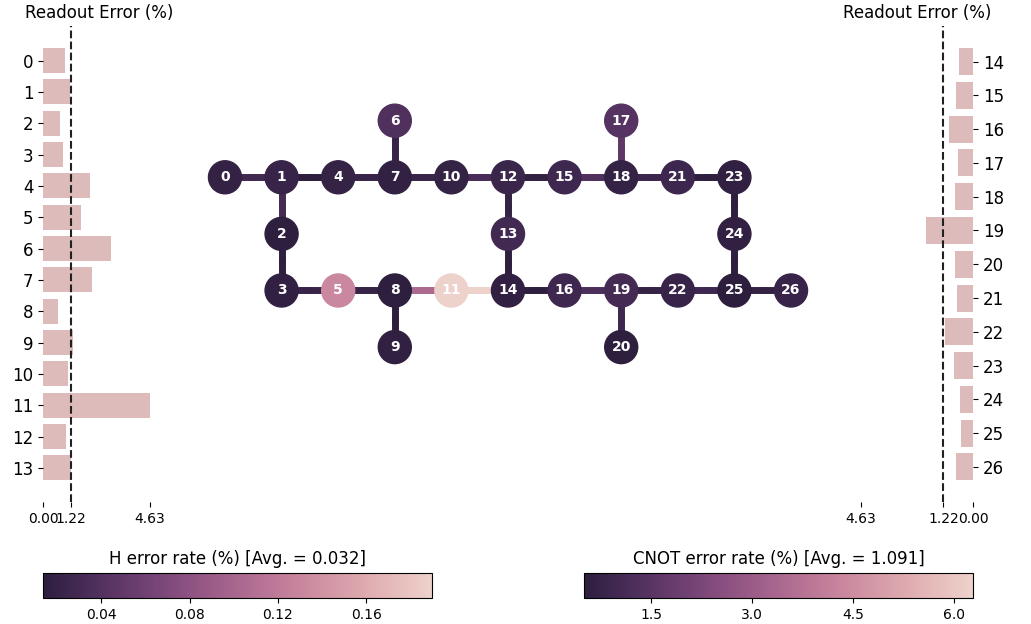}
    \caption{Error map for qubits and couplings of \texttt{imbq\_kolkata} device from 18th April, 2023.}
    \label{fig:230418}
\end{figure}

\begin{figure}[]
    \centering
    \includegraphics[width=0.48\textwidth]{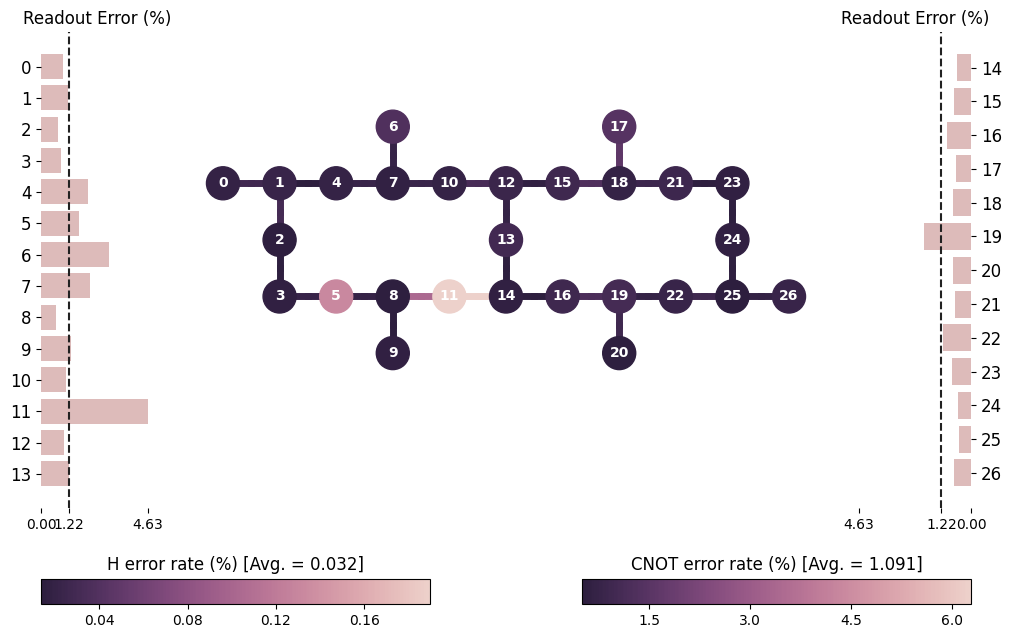}
    \caption{Error map for qubits and couplings of \texttt{imbq\_kolkata} device from 22nd April, 2023}
    \label{fig:230422}
\end{figure}

\begin{figure}[t]
    \centering
    \includegraphics[width=0.48\textwidth]{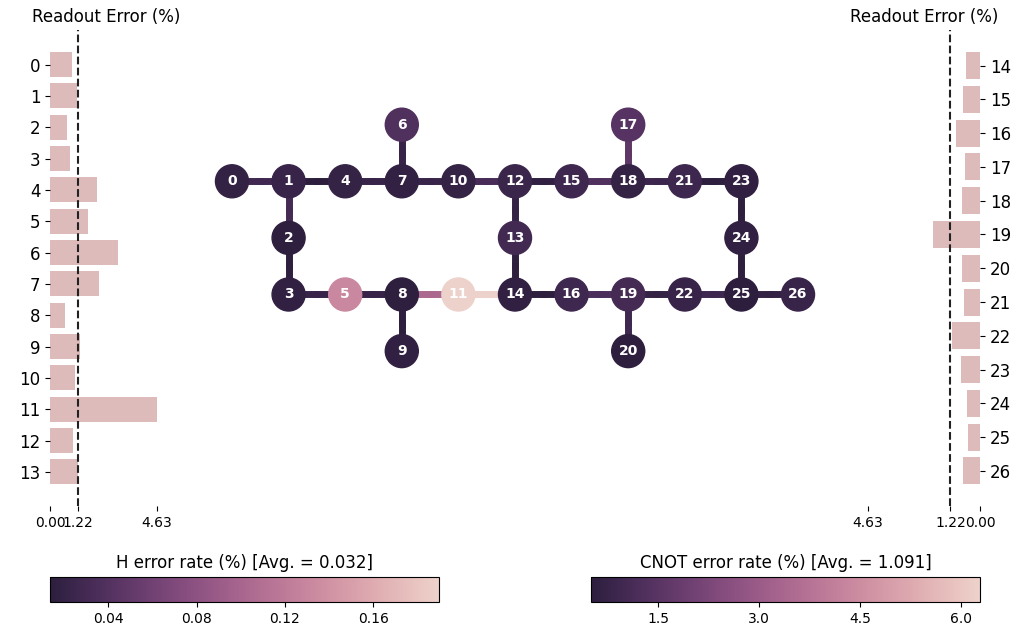}
    \caption{Error map for qubits and couplings of \texttt{imbq\_kolkata} device from 27th April, 2023}
    \label{fig:230427}
\end{figure}

\begin{figure}[]
    \centering
    \includegraphics[width=0.48\textwidth]{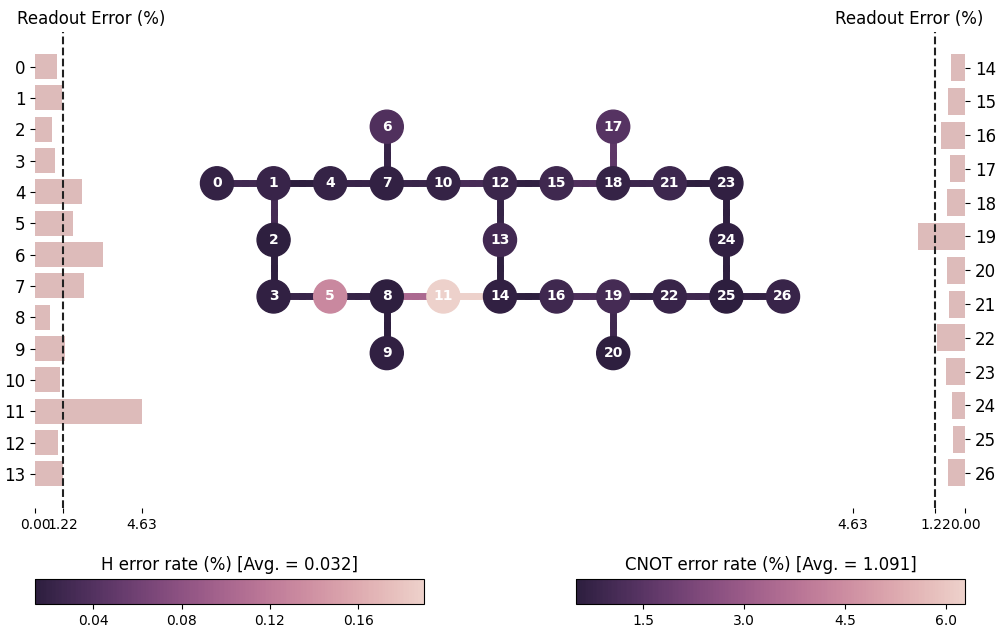}
    \caption{Error map for qubits and couplings of \texttt{imbq\_kolkata} device from 5th May, 2023}
    \label{fig:230505}
\end{figure}

The layout of \texttt{ibmq\_kolkata} device is shown in Fig.~\ref{fig:kolkata_layout}.
We use the 20 qubit loop consisting of qubits 1, 2, 3, 5, 8, 11, 14, 16, 19, 22, 25, 24, 23, 21, 18, 15, 12, 10, 7, and 4 for our demonstrations. We choose the link between qubit 12 and qubit 15 to place the twist defect. The readout length for all these runs was $640 ns$. $P_{01}$ ($P_{10}$) is the probability that measuring a qubit gives 0 (1) immediately after preparing it in state 1 (0). The readout error is indicated as RE, and the frequency $f$ and the anharmonicity $a$ are both in GHz.
The calibration data from 18th April, 2023 for the used qubits are given in 
Table~\ref{table:230418}, and the error map is shown in Fig.~\ref{fig:230418}. The calibration data from 22nd April, 2023 for the used qubits are given in Table~\ref{table:230422}, and the error map is shown in Fig.~\ref{fig:230422}. The calibration data from 27th April, 2023 for the used qubits are given in Table~\ref{table:230427}, and the error map is shown in Fig.~\ref{fig:230427}. The calibration data from 5th May, 2023 for the used qubits are given in Table~\ref{table:230505}, and the error map is shown in Fig.~\ref{fig:230505}. The error maps are generated using Qiskit's visualization library.
\begin{table}[]
\centering
\begin{tabular}{ |c|c| c| c| c| c| c| c| } 
 \hline
Qubit & T1($\mu$s) & T2($\mu$s) & $f$ & $a$& RE & $P_{01}$ & $P_{10}$  \\ 
 \hline
 1& 81.37 & 181.74 & 4.99 & $-0.34$ & 0.0087 & 0.0112 & 0.0062 \\
 \hline
 2&72.9 & 52.5 & 5.11 & $- 0.34$ & 0.0155 & 0.022 & 0.009 \\
 \hline
 3& 126.2 & 165.5 & 4.87 & $-0.35$ & 0.013 & 0.0122 & 0.0138 \\
 \hline
 5& 128.1 & 29.5 & 5.11 & $-0.342$ & 0.0325 & 0.0344 & 0.0306  \\
 \hline
 8& 135.2 & 60.4 & 4.93 & $-0.345$ & 0.0427 & 0.0292 & 0.0562  \\
 \hline
 11& 132.8 & 43.6 & 4.87 & $-0.373$ & 0.1247 & 0.1182 & 0.1312 \\
 \hline
 14& 58.82 & 70.08 & 5.12 & $-0.343$ & 0.06 & 0.0216 & 0.0984 \\
 \hline
 16& 78.78 & 116.12 & 5.22 & $-0.34$ & 0.0092 & 0.0078 & 0.0106 \\
 \hline
 19& 86.69 & 32.28 & 5.002 & $-0.345$ & 0.0383 & 0.0208 & 0.0558 \\
 \hline
 22&  108.02 & 39.49 & 5.127 & $-0.343$ & 0.0369 & 0.0352 & 0.0386 \\
 \hline
 25& 253.27 & 83.23 & 4.92 & $-0.347$ & 0.0098 & 0.0124 & 0.0072 \\
 \hline
 24&120.36 & 86.38 & 5.004 & $-0.346$ & 0.0092 & 0.0096 & 0.0088 \\
 \hline
 23&  153.89 & 48.45 & 5.138 & $-0.343$ & 0.0135 & 0.0214 & 0.0056 \\
 \hline
 21& 83.88 & 17.18 & 5.274 & $-0.341$ & 0.0073 & 0.0108 & 0.0038 \\
 \hline
 18& 163.04 & 104.59 & 5.097 & $-0.344$ & 0.0102 & 0.0126 & 0.0078 \\
 \hline
 15& 175.32 & 207.31 & 5.041 & $-0.344$ & 0.0053 & 0.007 & 0.0036 \\
 \hline
 12& 148.89 & 137.72 & 4.96 & $-0.346$ & 0.0088 & 0.01 & 0.0076 \\
 \hline
 10& 113.57 & 44.2 & 5.178 & $-0.342$ & 0.0151 & 0.014 & 0.0162  \\
 \hline
 7& 164.47 & 33.13 & 5.031 & $-0.346$ & 0.0223 & 0.0208 & 0.0238 \\
 \hline
 4& 87.74 & 87.38 & 5.224 & $-0.341$ & 0.0214 & 0.0224 & 0.0204 \\
 \hline
\end{tabular}
\caption{Calibration data for the qubits used for demonstrations on 18th April, 2023. }
\label{table:230418}
\end{table}

\begin{table}[]
\centering
\begin{tabular}{ |c|c| c| c| c| c| c| c| } 
 \hline
 Qubit & T1($\mu$s) & T2($\mu$s) & $f$ & $a$& RE & $P_{01}$ & $P_{10}$   \\ 
 \hline
  1& 17.686 & 106.2& 4.99& $-0.345$ & 0.0203 & 0.0272 & 0.0134 \\
 \hline
 2& 42.29& 37.02 & 5.117 & $-0.343$ & 0.0516 & 0.078 & 0.0252 \\
 \hline
 3& 143.82 & 188.71 & 4.866 & $-0.346$ & 0.0102 & 0.0114 & 0.009 \\
 \hline
 5& 103.63  & 30.41 & 5.113 & $-0.342$ & 0.032 & 0.0332 & 0.0308 \\
 \hline
 8& 163.56  & 50.83 & 4.928 & $-0.345$ & 0.0451 & 0.032 & 0.0582 \\
 \hline
 11& 139.18 & 39.37 & 4.868  & $-0.373$ & 0.1176 & 0.1116 & 0.1236  \\
 \hline
 14& 128.15 & 67.17 &5.117  & $-0.343$ & 0.0629  & 0.0246 & 0.1012 \\
 \hline
 16& 96.47 & 92.36 & 5.222 & $-0.34$ & 0.0111 & 0.0084 & 0.0138 \\
 \hline
 19& 112.28 & 26.24 & 5.002 & $-0.345$ & 0.0776 & 0.0316 & 0.1236  \\
 \hline
 22& 115.85  & 38.87 &5.127 & $-0.343$ & 0.0411 & 0.0383 & 0.0438 \\
 \hline
 25& 167.11 & 81.57  & 4.92 & $-0.347$ & 0.0088 & 0.0122 & 0.0054 \\
 \hline
 24& 122.28 & 83.58 & 5.004 & $-0.346$ & 0.0083 & 0.0108 & 0.0058 \\
 \hline
 23&  127.04 & 42.31 & 5.138 & $-0.343$ & 0.0143  & 0.026 & 0.0026 \\
 \hline
 21& 69.81 & 16.35 & 5.274  & $-0.341$ & 0.0069 & 0.0078 & 0.006 \\
 \hline
 18& 70.17 & 105.99 & 5.097 & $-0.344$ & 0.0087  & 0.0086 & 0.0088 \\
 \hline
 15& 157.17 & 192.35 & 5.041 & $-0.344$ & 0.0061 & 0.008 & 0.0042  \\
 \hline
 12& 144.32 & 174.84 & 4.961 & $-0.346$ & 0.0075  & 0.0092 & 0.0058 \\
 \hline
 10& 88.61 & 35.82& 5.178  & $-0.342$ & 0.0106 & 0.0144 & 0.0068 \\
 \hline
 7& 68.69 & 34.22 & 5.031 & $-0.345$ & 0.0242 & 0.0254 & 0.023 \\
 \hline
 4& 108.36 & 66.99 & 5.224 & $-0.341$ & 0.0216 & 0.0202 & 0.023 \\
\hline
 \end{tabular}
\caption{Calibration data for the qubits used for demonstrations on 22nd April, 2023.}
\label{table:230422}
\end{table}

\begin{table}[]
\centering
\begin{tabular}{ |c|c| c| c| c| c| c| c| } 
 \hline
 Qubit & T1($\mu$s) & T2($\mu$s) & $f$ & $a$& RE & $P_{01}$ & $P_{10}$    \\ 
 \hline
  1& 249.53 &176.45 & 4.991 & $-0.345$ & 0.0112 & 0.012 & 0.0098 \\
 \hline
 2& 10.43 & 44.90 & 5.117 & $-0.343$ & 0.0149& 0.019 & 0.0108 \\
 \hline
 3& 75.96 & 136.94 & 4.866 & $-0.346$ & 0.0143 & 0.0158 & 0.0128 \\
 \hline
 5& 133.18 & 31.47 & 5.113 & $-0.342$ & 0.0323 & 0.0324 & 0.0322  \\
 \hline
 8& 125.48 & 49.20 & 4.928 & $-0.345$ & 0.0342 & 0.0328& 0.0356  \\
 \hline
 11& 149.91 & 41.34 & 4.868 & $-0.373$ & 0.0944 & 0.0956 & 0.0932  \\
 \hline
 14& 165.77 & 204.37 & 5.117 & $-0.343$ & 0.0657 & 0.0244 & 0.107 \\
 \hline
 16& 106.17 & 84.22 & 5.222 & $-0.34$ & 0.009 & 0.007& 0.011   \\
 \hline
 19& 54.53 & 23.02 & 5.002 & $-0.345$ & 0.0387 & 0.0252 & 0.0522 \\
 \hline
 22& 138.54 & 34.42& 5.127 & $-0.343$ & 0.0372 & 0.037 & 0.0374 \\
 \hline
 25& 222.14 & 120.88 & 4.921 & $-0.347$ & 0.0109 & 0.0156 & 0.0062 \\
 \hline
 24& 93.44 & 106.77 & 5.005 & $-0.346$ & 0.0094 & 0.0104 & 0.0084 \\
 \hline
 23& 135.32 & 52.53 & 5.138 & $-0.343$ & 0.0035 & 0.005 & 0.002 \\
 \hline
 21& 124.1 & 18.48 & 5.274 & $-0.341$ & 0.0047 & 0.0052 & 0.0042 \\
 \hline
 18& 117.51 & 148.1 & 5.097 & $-0.344$ & 0.0094 & 0.0112 & 0.0076 \\
 \hline
 15& 129.05 & 224.34 & 5.041 & $-0.344$ & 0.0067 & 0.0072 & 0.0064 \\
 \hline
 12& 205.67 & 225.1 & 4.961 & $-0.346$ & 0.0075 & 0.0084 & 0.0066 \\
 \hline
 10& 85.16 & 38.17 &  5.178 & $-0.342$ & $0.0087$ & 0.0094 & 0.008  \\
 \hline
 7& 170.34 & 45.36 & 5.031 &$-0.346$ & 0.0261 & 0.0202 & 0.032 \\
 \hline
 4& 122.95 & 100.49 & 5.224 & $-0.341$& 0.0196 & 0.021 & 0.0182 \\
\hline
 \end{tabular}
\caption{Calibration data for the qubits used for demonstrations on 27th April, 2023. }
\label{table:230427}
\end{table}

\begin{table}[]
\centering
\begin{tabular}{ |c|c| c| c| c| c| c| c| } 
 \hline
 Qubit & T1($\mu$s) & T2($\mu$s) & $f$ & $a$& RE & $P_{01}$ & $P_{10}$   \\ 
 \hline
  1& 176.13 & 214.04 & 4.991 & $-0.345$ & 0.0099 &0.0146& 0.0052  \\
 \hline
 2& 10.9 & 44.9 & 5.113 & $-0.343$ & 0.0149 & 0.019 & 0.0108 \\
 \hline
 3& 96.79 & 195.85 & 4.866 & $-0.346$ & 0.0128 & 0.0118 & 0.0138 \\
 \hline
 5& 88.86 & 29.91 & 5.113 & $-0.342$ & 0.0348 & 0.0348 & 0.0348  \\
 \hline
 8& 129.58 & 52.25 & 4.928 & $-0.345$ & 0.0277 & 0.0248 & 0.0306  \\
 \hline
 11& 133.31 & 37.53 & 4.868 & $-0.373$ & 0.123 & 0.1296 & 0.1164   \\
 \hline
 14& 156.54 & 169.92 & 5.117  & $-0.343$ & $0.066$ & 0.022 & 0.11 \\
 \hline
 16& 113.15 & 103.99 & 5.222 & $-0.3402$ & 0.0079 & 0.008 & 0.0078  \\
 \hline
 19& 105.14 & 28.67 & 5.002 & $-0.345$ & 0.0699 & 0.025 & 0.1148   \\
 \hline
 22& 116.19 & 32.82 & 5.127 & $-0.343$ & 0.0382 & 0.0344 & 0.042 \\
 \hline
 25& 194.42 & 73.42 & 4.921 & $-0.347$ & 0.0056 & 0.0072 & 0.004 \\
 \hline
 24& 107.53 & 107.29 & 5.005 & $-0.346$ & 0.0097 & 0.0106 & 0.0088 \\
 \hline
 23& 115.88 & 50.13 & 5.138 & $-0.343$ & 0.0044 & 0.0058 & 0.003  \\
 \hline
 21& 94.18 & 19.34 & 5.274 & $-0.341$ & 0.0062 & 0.0062 & 0.0062 \\
 \hline
 18& 145.87 & 125.4 & 5.097 & $-0.344$ & 0.0092 & 0.0104 & 0.008 \\
 \hline
 15& 91.18 & 153.23 & 5.041 & $-0.344$ & 0.0066 & 0.0084 & 0.0048  \\
 \hline
 12& 175.97 & 188.79 & 4.961 & $-0.346$ & 0.0078 & 0.009 & 0.0066  \\
 \hline
 10& 120.79 & 38.97 & 5.178 & $-0.342$ & 0.015 & 0.0092 & 0.0208 \\
 \hline
 7& 144.69 & 54.6 & 5.031 & $-0.346$ & 0.0239 & 0.0292 & 0.0186  \\
 \hline
 4& 145.79 & 139.71 & 5.225 & $-0.341$ & $0.0232$ & 0.0248 & 0.0216 \\
\hline
 \end{tabular}
\caption{Calibration data for the qubits used for demonstrations on 5th May, 2023. }
\label{table:230505}
\end{table}

\end{document}